\documentclass[journal]{IEEEtran}

\usepackage{xcolor,soul,framed} %,caption

\colorlet{shadecolor}{yellow}
\usepackage[pdftex]{graphicx}
\graphicspath{{../pdf/}{../jpeg/}}
\DeclareGraphicsExtensions{.pdf,.jpeg,.png}

\usepackage[cmex10]{amsmath}
\usepackage{array}
\usepackage{mdwmath}
\usepackage{mdwtab}
\usepackage{eqparbox}
\usepackage{url}

\begin{document}
\bstctlcite{IEEEexample:BSTcontrol}
    \title{An Iterative Least Squares Method for Proton CT Image Reconstruction.}
  \author{Don~F.~DeJongh and
      Ethan~A.~DeJongh
  \thanks{Research reported in this publication was supported by the National Cancer Institute of the National Institutes of Health under award number R44CA243939.}
  \thanks{© 2021 IEEE.  Personal use of this material is permitted.  Permission from IEEE must be obtained for all other uses, in any current or future media, including reprinting/republishing this material for advertising or promotional purposes, creating new collective works, for resale or redistribution to servers or lists, or reuse of any copyrighted component of this work in other works.}
  \thanks{Don F. DeJongh is with ProtonVDA LLC, 1700 Park St Ste 208, Naperville, IL 60563 USA (e-mail: fritz.dejongh@protonvda.com).}% <-this % stops a space
  \thanks{Ethan A. DeJongh is with ProtonVDA LLC, 1700 Park St Ste 208, Naperville, IL 60563 USA (e-mail: ethan.dejongh@protonvda.com).}% <-this % stops a space
}

% The paper headers
%\markboth{IEEE Transactions on Radiation and Plasma Medical Sciences}{}

% ====================================================================
\maketitle

% === ABSTRACT ====================================================================
% =================================================================================
\begin{abstract}
%\boldmath
Clinically useful proton Computed Tomography images will rely on algorithms to find the three-dimensional proton stopping power distribution that optimally fits the measured proton data.  We present a least squares iterative method with many features to put proton imaging into a more quantitative framework.  These include the definition of a unique solution that optimally fits the protons, the definition of an iteration vector that takes into account proton measurement uncertainties, the definition of an optimal step size for each iteration individually, the ability to simultaneously optimize the step sizes of many iterations, the ability to divide the proton data into arbitrary numbers of blocks for parallel processing and use of graphical processing units, and the definition of stopping criteria to determine when to stop iterating.  We find that it is possible, for any object being imaged, to provide assurance that the image is quantifiably close to an optimal solution, and the optimization of step sizes reduces the total number of iterations required for convergence. We demonstrate the use of these algorithms on real data.
\end{abstract}

% === KEYWORDS ====================================================================
% =================================================================================
\begin{IEEEkeywords}
proton imaging, proton computed tomography, least squares, iterative algorithm, relaxation coefficient, parallel processing, stopping criterion
\end{IEEEkeywords}

% For peer review papers, you can put extra information on the cover
% page as needed:
% \ifCLASSOPTIONpeerreview
% \begin{center} \bfseries EDICS Category: 3-BBND \end{center}
% \fi
%
% For peerreview papers, this IEEEtran command inserts a page break and
% creates the second title. It will be ignored for other modes.
\IEEEpeerreviewmaketitle

% ====================================================================
% ====================================================================
% ====================================================================

% === I. INTRODUCTION =============================================================
% =================================================================================
\section{Introduction}

\IEEEPARstart{I}{n} radiation therapy, protons provide a superior dose distribution compared to x rays, with a relatively low dose deposition in the entrance region (plateau), followed by a steep increase to a dose (Bragg) peak and an even steeper distal dose fall-off~\cite{Scholz}. {The steep distal dose gradient and finite range of the protons necessitate accurate knowledge of the range in the patient.} One source of range uncertainty is
the use of x-ray imaging for treatment planning to obtain a map of relative stopping power (RSP) of tissues (relative to water), which is inaccurate due to the differences in the dependence of x-ray attenuation and proton energy loss on tissue composition (electron density and atomic number). This yields an inherently inaccurate conversion of x-ray Hounsfield units to proton RSP.

Treatment planning procedures take these uncertainties into account with measures including adding uncertainty margins, selection of beam angles tangential to organs at risk, and robust optimization.  Using dose delivery technology such as Pencil Beam Scanning (PBS) and intensity modulation, the resulting plans are robust to the uncertainties and provide major benefits to a significant fraction of patients \cite{Lomax}.  However, they increase the high-dose treatment volume and can preclude use of the most advantageous beam angles.

In the quest to further optimize proton therapy while also reducing costs, proton beam-based image guidance is often considered to be a prerequisite to achieve the full potential of proton therapy \cite{Schreuder}. This is particularly the case for hypo-fractionated treatments, which can benefit from more conformal dose distributions and a higher standard of safety given the high dose delivery for each treatment. Proton radiography (pRad) has the potential to provide a fast and efficient check of patient set up and integrated range along a beam’s eye view just before treatment \cite{Miller}\cite{Ordonez}\cite{Pankuch}. {Proton CT (pCT) can reduce range uncertainties} and substantially reduce the uncertainties of treatment planning by directly measuring RSP without being affected by image artifacts and with much lower dose to the patient than comparable x-ray images \cite{Schulte}.

\begin{figure}
  \begin{center}
  \includegraphics[width=\columnwidth]{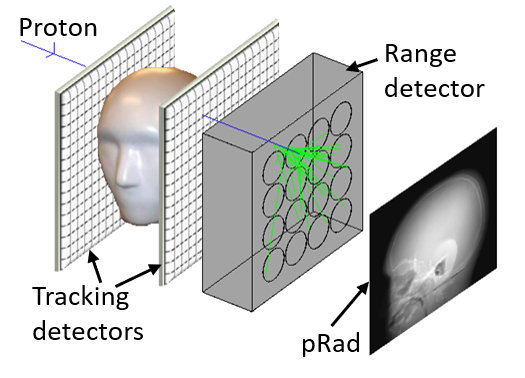}\\
  \caption{Illustration of proton imaging, with tracking and residual range measurements for each proton. The proton radiograph image on the right displays range through the patient vs. transverse position.}\label{system}
  \end{center}
\end{figure}

Proton imaging uses tracking detectors to measure the transverse positions of individual protons before and after the patient, and a residual range detector to determine the proton energy absorbed within the patient, as illustrated in Fig.~\ref{system}. A two-dimensional pRad image uses a single projection angle, directly quantifying proton range through the patient rather than integrated x-ray attenuation. A three-dimensional pCT image measures the 3D RSP map of the patient by acquiring proton histories from a full set of projection angles. Proton trajectories deviate from straight lines due to multiple Coulomb scattering, thus blurring images. {Iterative reconstruction algorithms} \cite{Giacometti} {use, for each proton, an estimate of its most likely path, along with its energy loss quantified as water-equivalent path length (WEPL), to obtain images with improved spatial resolution.}  Another approach uses distance-driven binning with filtered backprojection to account for the curved trajectories and reconstruct the image \cite{Rit}.

\begin{figure}
  \begin{center}
  \includegraphics[width=0.75\columnwidth]{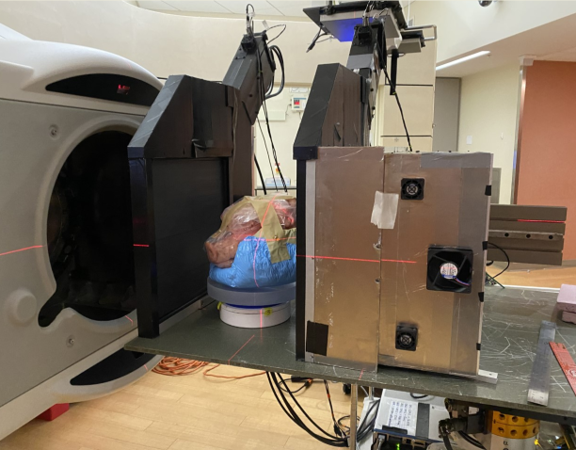}
  \includegraphics[width=0.2\columnwidth]{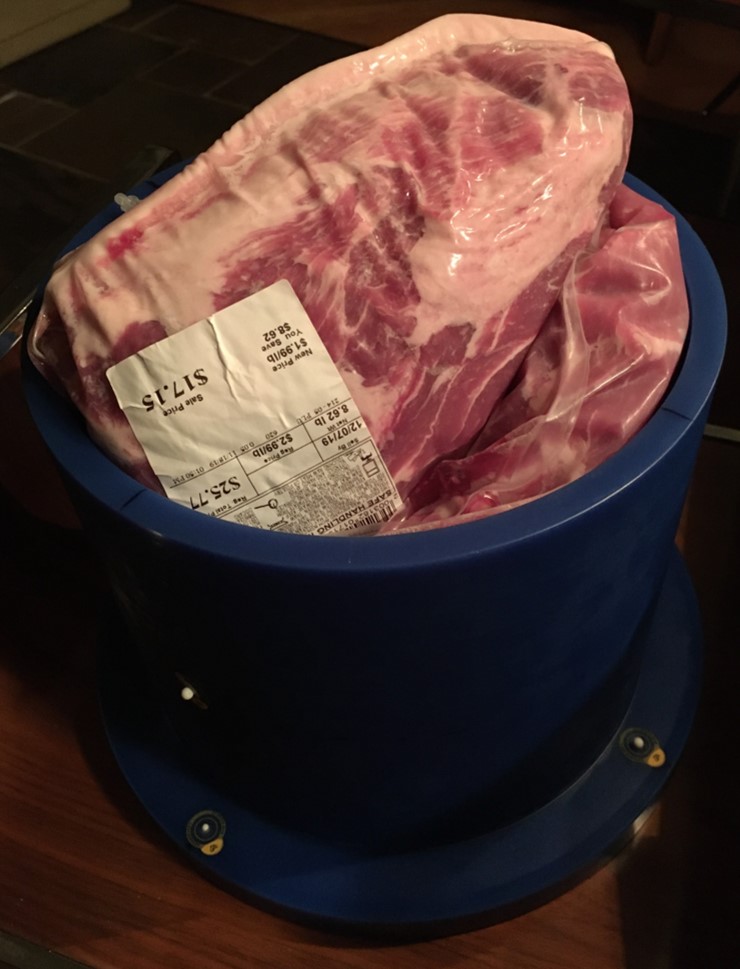}\\
  \caption{The ProtonVDA system \cite{dejongh2020technical}, {positioned in the horizontal beam treatment room at the NMCPC. The rotating platform between the tracking planes enabled imaging with a full set of angles relative to the PBS system, and was used with a pig's head (as shown) as well as with the pork shoulder and ribs shown on the right.}}\label{setup}
  \end{center}
\end{figure}

%\begin{figure}
%  \begin{center}
%  \includegraphics[width=\columnwidth]{figs/pRad.png}\\
%  \caption{Left: ProtonVDA system \cite{dejongh2020technical} with pediatric head phantom placed between tracking planes, prepared to take data %with pencil beam scanning. Right: The resulting proton radiograph \cite{Ordonez}\cite{sarosiek2020prototype} was automatically and promptly %produced, and displays accurate water equivalent thickness.}\label{pRad}
%  \end{center}
%\end{figure}

\begin{figure}
  \begin{center}
  \includegraphics[width=\columnwidth]{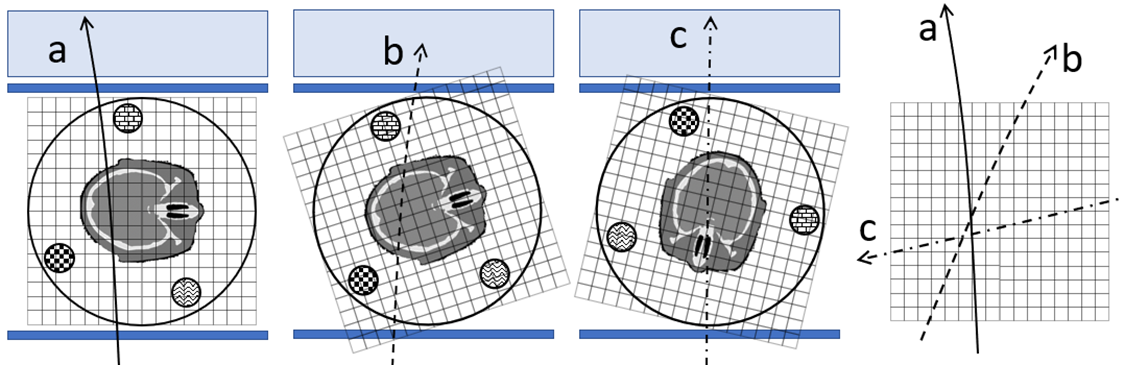}\\
  \caption{Protons a, b, and c are acquired with the object at different angles relative to the detectors.  For image reconstruction, all proton trajectories are placed in a single 3D coordinate grid, as shown on the right, that moves with the object.}\label{pCT}
  \end{center}
\end{figure}

The first challenge in producing clinically useful pCT images is to efficiently obtain a large data set of protons with accurately measured trajectories and well-calibrated WEPLs~\cite{Johnson_2017}.  We acquired the data for the images presented herein with the ProtonVDA proton imaging system~\cite{dejongh2020technical}, which is based on well-established fast-scintillator technology, and is fast, compact, monolithic, and easily scaled to large field sizes (40 x 40 cm$^2$ in the current implementation).  The ProtonVDA system (Fig.~\ref{setup}) is able to automatically and promptly produce pRad images~\cite{sarosiek2020prototype}, and has also produced our first pCT images~\cite{dejongh2020comparison} which required the measurement of protons through the object at a comprehensive set of angles and the placement of proton trajectories into a single 3D coordinate grid for image reconstruction, as illustrated in Fig.~\ref{pCT}. Our standard voxel size is 1 mm$^3$, which is matched to our expected spatial resolution.
{The data shown in Fig.}~\ref{pork} {is for the sample of pork shoulder and ribs inserted into a blue wax cylinder shown in Fig.}~\ref{setup}. {We acquired this data at the Northwestern Medicine Chicago Proton Center (NMCPC) using protons of four different energies, 120, 160, 185, and 203 MeV, delivered with a pencil beam scanning system across the volume at 90 projection angles spaced 4 degrees apart. Approximately 180 million protons were used to reconstruct a 3-D volume of $250 \times 250 \times 250$ 1 mm$^3$ voxels.  Similarly, the pCT image of the phantom shown in Fig.}~\ref{george} {used approximately 20 million protons taken at incoming energies of 118, 160, and 187 MeV to reconstruct a volume of $200 \times 60 \times 200$ 1 mm$^3$ voxels.}

The second challenge in producing clinically useful pCT images is to reconstruct the 3D RSP distribution that optimally fits the proton data by solving the following matrix equation for $x$:
\begin{equation}
    Ax = b,
\end{equation}
where $b$ is a vector with one entry per proton, containing the WEPL measurements for each proton, $x$ is a vector with one entry per voxel, containing the RSP for that voxel, and $A$ is a matrix with one row for each proton and one column for each voxel, where each entry contains the chord length of the proton trajectory, {as measured by the tracking detectors}, through the voxel.  Since each proton touches only a tiny fraction of the voxels, $A$ is quite sparse.

Penfold and Censor have described several iterative algorithms which adjust the RSPs of the voxels touched by the protons to match the WEPLs of the protons \cite{Penfold}. These algorithms generally rely on a projective approach, with repeated projections of a solution vector $x_k$ onto hyperplanes in a space with coordinates defined as the components of $x$. Each hyperplane is defined by an equation obtained from one row of $A$ multiplied by $x$. The goal of the projections is to move towards a solution consistent with each proton, and these methods have been successfully applied to several pCT data sets.  These algorithms are often combined with additional smoothing algorithms such as median filtering or total variation superiorization \cite{Schultze}. 

Goitein described in 1972 an iterative least squares algorithm used for reconstructing the first tomographic images using charged particles, with a formalism able to accommodate tracks with curved trajectories, and a prescription for optimizing the step size of each iteration \cite{Goitein}.  
{The least squares approach is suitable for pCT imaging, which utilizes measurements of many protons with large WEPL uncertainties.  A typical proton going through 200 mm of material will have a WEPL uncertainty of around 3 mm, arising from range straggling effects as well as from the precision of the range detector.  In contrast, the WEPL uncertainty of a path through a 1 mm voxel with RSP known to 1\% is roughly a factor of 300 smaller. In this case, the precision of the RSP measurement is arrived at through the averaging effect of roughly $10^5$ protons touching each voxel.} The goal of each iteration is to converge towards a point that is a best fit but that is not perfectly consistent with each proton.  {A similar approach uses a penalized least squares determination with a noise suppressing roughness penalty}~\cite{Hansen_2016}. 

%\begin{figure}
%  \begin{center}
%  \includegraphics[width=3.5in]{figs/projective.png}\\
%  \caption{Illustration \cite{Penfold} of the projective approach to a solution to Ax = b for the case of two voxels, where the %coordinates in the plane represent possible solutions for the two entries in x.  The lines represent hyperplanes, each defined %by an equation obtained from one row of A multiplied by x.  The case shown is noiseless, and is converging to a solution %consistent for each proton.}\label{projective}
%  \end{center}
%\end{figure}

% =================================================================================

{Previous studies have generally analyzed reconstructed pCT images with respect to a known ground truth based on simulated data, or real data from phantoms constructed from a number of known materials. We see a need for methods with metrics that are applicable to general unknown objects being imaged.} Our goal is to produce a proton imaging system with prompt image reconstruction, using a method for which each iteration is as fast as possible, each iteration is as useful as possible (moves as much as possible towards an optimal solution), and that iterates only as many times as necessary (stops when close enough to the optimal solution).  We present herein an iterative least squares method for pCT image reconstruction that achieves this with many features to put pCT imaging into a more quantitative framework.  These include:
\begin{itemize}
\item	{The least squares formalism defines} a unique solution that optimally fits the protons.  Effectively converging toward this solution eliminates some problems often associated with projective algorithms \cite{Penfold2}:
\begin{itemize}	
\item The solution does not depend on the initial starting point for the iterations.  
\item	There is no need to have a trade-off between optimizing spatial resolution and RSP resolution.
\end{itemize}
\item	{The least squares formalism leads to an iteration vector that takes into account the large WEPL uncertainties in the proton data, in contrast to projection methods based on the assumption that all hyperplanes intersect at a unique point.}
\item	The ability to optimize the step size for each iteration individually.
\item	The ability to simultaneously optimize the step sizes of many iterations.
\item The ability to divide the proton data into arbitrary numbers of blocks.  Blocks can be as small as a single proton and still maintain the “simultaneous” character of the algorithm, which takes into account all protons on an equal basis for each iteration, rather than taking them into account “sequentially,” in which case the last proton has a disproportionate impact.
\begin{itemize}	
\item	The resulting flexibility is very useful for optimizing use of computing resources such as {GPUs}~\cite{Hansen_2014}{.}
\end{itemize}
\item	The definition of stopping criteria, to determine when to stop iterating.
\begin{itemize}	
\item	As a result, it is possible, for any object being imaged, to provide assurance that the image is quantifiably close to an optimal solution.
\end{itemize}
\end{itemize}

{Using this framework, quantifying how close the reconstructed image is to the solution that optimally fits the data does not depend on knowledge of a ground truth.} The assurance of an image reconstruction quantifiably close to an optimal solution is a crucial step towards applying this technology to clinical treatment planning, and a useful starting point for evaluating the clinical impact of further image processing or use of approximations.  

We have not incorporated smoothing methods, seeing these as better left for a later step after defining an optimal solution. While these can produce better-looking images with less noise, they can introduce unknown systematic effects, particularly when imaging complex objects with rapid density variations.

While we are describing a system for proton imaging, the methods apply equally as well to other ions such as helium and can also be applied to other tomographic modalities such as x-ray imaging.

\section{Solving $Ax = b$ using an iterative least squares method}

A typical pCT image may reconstruct a few million voxels using a few hundred million protons, each with a WEPL measurement with approximately 3 mm uncertainty.  The $A$ matrix is therefore “tall and skinny” with no solution that exactly fits all protons.  We define the proton deviation vector as:
\begin{equation}\label{dp}
   d_p  = A x - b 
\end{equation}
Here, $d_{p}^i$ is the deviation for proton $i$, with $d_p \neq 0$ even for the best solution.
We then define the voxel deviation vector $d_v$ as a weighted average of the $d_p$ of all the protons going through each voxel, with $d_{v}^j$ as the deviation for voxel $j$.
Each iteration updates the voxels using the voxel deviation vector:
\begin{equation}\label{iterate}
    x \rightarrow x - \lambda d_v
\end{equation}
where $\lambda$ is a relaxation coefficient that determines the step size of the iteration and can vary with iteration.

Our current choice for the weighted average is to use the chord lengths for the weights, where the deviation for a voxel can be written as 
\begin{equation}
    \frac{\sum_i \text{a}_i d_p^i}{\sum_i \text{a}_i}, 
\end{equation}
where the sums are over all protons touching the voxel and the $\text{a}_i$ are the chord lengths for each proton.  The weights could possibly be further optimized, for example, by incorporating the individual precision for the WEPL measurement of each {proton} \cite{Goitein}\cite{Hansen_2016}{.}  For our application, the protons are all measured with {approximately similar precision, since the detector uncertainty dominates the WEPL uncertainty at small WEPL, and adds in quadrature to the range straggling which tends to dominate at higher WEPL. We have made the approximation that the WEPL uncertainty is the same for each proton.} For simplicity we have not explicitly incorporated this uncertainty into the following equations.

In terms of the $A$ matrix, we can write the voxel deviation vector as:
\begin{equation}
d_{v}^j = \frac{(A^T d_p)_j}{\sum_i \alpha^T_{ji}} 			
       = \frac{A^T_j}{\sum_i \alpha^T_{ji}} d_p				
\end{equation}
where $\alpha^T_{ji}$ are elements of $A^T_j$, which is the $j$th row of $A^T$.  We define the $\bar{A}^T$ matrix as:
\begin{equation}\label{abar}
\bar{A}^T_j = \frac{A^T_j}{\sum_i \alpha^T_{ji}}		
\end{equation}
in terms of which we can write:
\begin{equation}\label{dv}
d_v = \bar{A}^T d_p.
\end{equation}

Our method is an example of a general Landweber iterative method \cite{Han}, for which broad convergence conditions have been established, with 
\begin{equation}
\bar{A}^T = V^{-1}A^T 
\end{equation}
\begin{equation}
V^{-1} = \text{diag}(1/ \sum_i \alpha^T_{ji} ) .
\end{equation}

We define $\chi^2$ as:
\begin{equation}
\chi^2 = d_p \cdot d_p					 
    =  (Ax - b) \cdot (Ax - b) 			
\end{equation}
\begin{equation}
\frac{\partial\chi^2}{\partial x_j} = 2 A^T_j \cdot (Ax) - 2 A^T_j \cdot b
\end{equation}
and $\partial\chi^2 / \partial x_j$ corresponds to the gradient used in Landweber iteration.  To obtain the least squares solution we set $\partial\chi^2 / \partial x_j = 0$, divide by $\sum_i \alpha^T_{ji}$, and apply (\ref{abar})  to obtain:
\begin{equation}
\bar{A}^TAx - \bar{A}^T b = 0
\end{equation}
Applying (\ref{dp}) and (\ref{dv}), this is equivalent to:
\begin{equation}
d_v = 0					
\end{equation}
Thus, we see the iteration in (\ref{iterate}) converges towards the unique least squares solution that optimizes the fit of the final image to the proton data.  Our goal with the use of a weighted average in the definition of $d_v$ is to obtain an optimal direction for the iteration vector, but it is also possible to converge while defining $d_v = A^T d_p$ if this proves to have a computational or numerical advantage.

\section{Choice of Relaxation Coefficient}\label{relaxation}

The steps to execute for iteration $k+1$ would most obviously be written as:
\begin{gather}
    x_{k+1} = x_{k} - \lambda_k d_{vk}\label{step1}	\\
d_{p(k+1)} = Ax_{k+1} - b\label{step2} 	\\		
d_{v(k+1)} = \bar{A}^T d_{p(k+1)\label{step3}}  		
\end{gather}
These steps require a choice of $\lambda_k$ before executing the computationally costly matrix-vector multiplications in (\ref{step2}) and (\ref{step3}). By substituting the value of $x_{k+1}$ from (\ref{step1}) into (\ref{step2}) and then (\ref{step2}) into (\ref{step3}), we re-write the last two steps as:
\begin{gather}
d_{p(k+1)} = d_{pk} - \lambda_k Ad_{vk}\label{advstep} \\
d_{v(k+1)} = d_{vk} - \lambda_k \bar{A}^T(Ad_{vk})\label{abartstep}	
\end{gather}
In this form it is  possible to execute the computationally costly matrix-vector multiplications $Ad_{vk}$  followed by $\bar{A}^T(Ad_{vk})$ before choosing a value for the relaxation coefficient, and furthermore, to utilize the resulting vectors in the choice of $\lambda_k$.  Some choices we find useful include:
\begin{itemize}
    \item Minimize $\chi^2_{k+1}$. The following expression was previously derived by Goitein \cite{Goitein}.
\begin{equation}
\begin{split}
\chi^2_{k+1} &= d_{p(k+1)} \cdot d_{p(k+1)}	\\
              &= \chi^2_k - 2\lambda_k d_{pk} \cdot (Ad_{vk}) + \chi^2_k |Ad_{vk}|^2	
\end{split}
\end{equation}
\begin{equation}
\frac{\text{d}\chi^2_{k+1}}{\text{d}\lambda_k} =  - 2 d_{pk} \cdot (Ad_{vk}) + 2\lambda_k |Ad_{vk}|^2  =  0	
\end{equation}
\begin{equation}\label{opt-lambda}
\lambda_k = \frac{d_{pk} \cdot (Ad_{vk})}{ |Ad_{vk}|^2} 				
\end{equation}
\item Make $\sum_i d_{v(k+1)}^i = 0$. 
\begin{equation}\label{center-lambda}
\lambda_k = \frac{\sum d_{vk}}{\sum \bar{A}^T(Ad_{vk})}				
\end{equation}
\item Minimize $d_{v(k+1)} \cdot d_{v(k+1)}$
\begin{equation}\label{dv-lambda}
\lambda_k = \frac{d_{vk} \cdot \bar{A}^T(Ad_{vk})}{ |\bar{A}^T(Ad_{vk})|^2} 			
\end{equation}
\end{itemize}

Thus, we find the interesting result that an optimal step size for each iteration, when looked at individually, can be applied, with (\ref{opt-lambda}).  Our experience is that the optimal relaxation coefficient can vary over two orders of magnitude from iteration to iteration, and the traditional method of choosing a constant $\lambda_k$ can be quite ineffectual.  If most voxels have $d_v$ far from 0, a smaller $\lambda_k$ is required, since each proton will be affected by many voxels.  If only a small number of voxels have $d_v$ far from 0, a larger $\lambda_k$ is possible.  In this situation, a constant $\lambda_k$ will result in a very gradual movement toward the optimal solution.  We have observed that it is often beneficial to use (\ref{center-lambda}) or (\ref{dv-lambda}), especially when $d_{v(k+1)}$ departs significantly from 0.  While less then optimal for the current step, this often provides conditions for subsequent large steps.  Various strategies are possible to combine different methods of choosing $\lambda$ at different iterations, as illustrated in Fig.~\ref{chi2p_alt}.

\begin{figure}
  \begin{center}
  \includegraphics[width=\columnwidth]{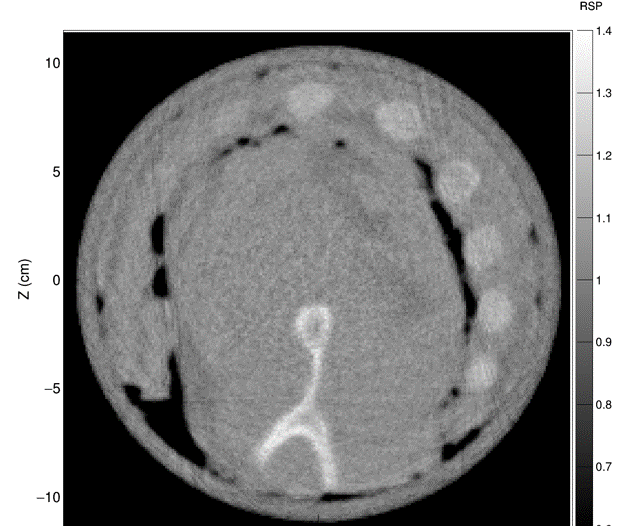}\\
    \includegraphics[width=\columnwidth]{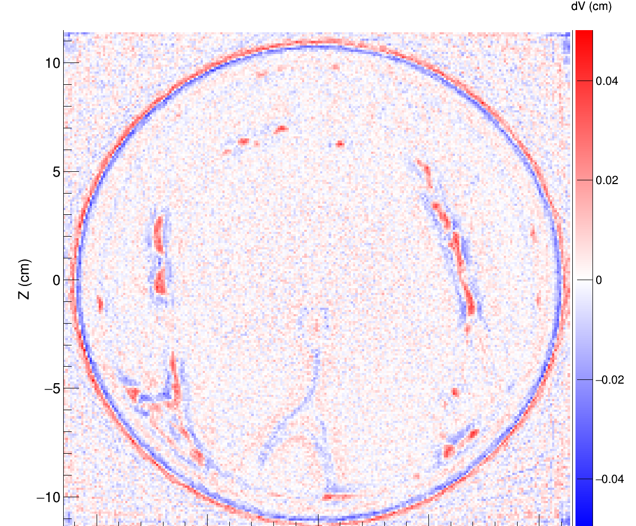}\\
  \includegraphics[width=\columnwidth]{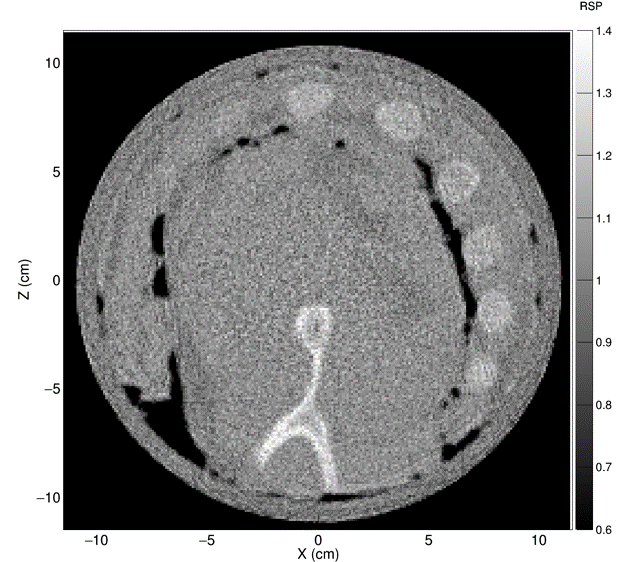}
\caption{A 1 mm thick pCT slice of the pork shoulder and ribs in Fig.~\ref{setup} \cite{dejongh2020comparison}.  
Top:  {Image from an early iteration.}
Middle:  $d_v$ for voxels in the top image, with large deviations mainly near edges. A large $\lambda$ was prescribed to go to the next iteration.
Bottom:  Final image, for which $d_v$ was low everywhere.
}\label{pork}
  \end{center}
\end{figure}

An example is shown in Fig.~\ref{pork} for the sample of pork shoulder and ribs in Fig.~\ref{setup}.  We have found that after a few iterations the largest deviations are usually around the edges, and use of (\ref{opt-lambda}) can enable a subsequent large step that results in images sharper than efficiently attainable with previous approaches, as seen with the bottom image of Fig.~\ref{pork}, which is quantified as being very close to the optimal solution.  The measured RSPs in uniform regions-of-interest agreed well with those derived from an x-ray CT image with a standard conversion of Hounsfield units to RSP \cite{dejongh2020comparison}.

\section{Global Optimization of Many Iterative Steps}

The idea in (\ref{advstep}) and (\ref{abartstep}) can be generalized to defer choice of relaxation coefficients for an arbitrary number of steps, and the combination of these steps can be globally optimized.  We define the $p$ and $v$ vectors as follows:
\begin{gather}
p_0 = d_{p0}\label{p0def}\\
v_k = \bar{A}^T p_k\label{vdef}\\
p_{k+1} = Av_k\label{pdef}		
\end{gather}
It is evident by induction using (\ref{p0def}) to (\ref{pdef}) and the definitions of $d_p$ and $d_v$ that the following can be written as a sum of the $p$ and $v$ vectors with coefficients $\kappa_i$ for a given number of iterations $n$:
\begin{gather}
d_{pn} = p_0 + \sum_{i=1}^n \kappa_i p_i\label{dpn}\\
d_{vn} = v_0 + \sum_{i=1}^n \kappa_i v_i\label{dvn}
\end{gather}
The solution vector $x$ can then be written as:
\begin{equation}\label{xn}
    x_n = x_0 + \sum_{i=1}^n \kappa_i v_{i-1}			
\end{equation}
as can be easily verified by substituting (\ref{xn}) into (\ref{dp}), applying (\ref{pdef}), and comparing with (\ref{dpn}).

The $\chi^2$ after $n$ iterations is, with $\kappa_0 := 1$:
\begin{align}
\chi^2_p &= d_{pn} \cdot d_{pn}\\
    &= \left(\sum_{i=0}^n \kappa_i p_i \right) \cdot \left(\sum_{i=0}^n \kappa_i p_i\right)\\
    &= \sum_{i,j=0}^n \kappa_i \kappa_j\  p_i \cdot p_j
\end{align}
After minimizing $\chi^2$ with respect to the $\kappa_i$, we can find the $x_n$ closest to the optimum solution using (\ref{xn}) with no need for the $\lambda_k$.  {(In principle, $\lambda_k$ values can be derived from the $\kappa_i$, and we have found in general that they are complex numbers.)} One direct way of finding the minimum is to set the partial derivatives of $\chi^2$ with respect to the $\kappa_i$ to zero to obtain:
\begin{equation}
    p_i \cdot p_0 + \sum_{j=1}^n \kappa_j\ p_i \cdot p_j  = 0
\end{equation}
Defining $P^n$ as the array of $p_i \cdot p_j$, $k^n$ as the vector of $\kappa_i$, and $p^n$ as the vector of $-p_i \cdot p_0$, the problem reduces to solving for $k^n$ in the following equation, for which there are many standard methods.
\begin{equation}\label{dpkn}
    P^n k^n = p^n
\end{equation}
Alternatively, $\chi^2$ can be defined from the $d_{vn}$ and since as described above $d_v = 0$ for the optimal result, the following, with similar definitions, leads to a similar solution as (\ref{dpkn}):
\begin{gather}
    \chi^2_v = d_{vn} \cdot d_{vn}\label{chi2dv}\\
    V^n k^n = v^n\label{dvkn}
\end{gather}

\begin{figure}
  \begin{center}
  \includegraphics[width=\columnwidth]{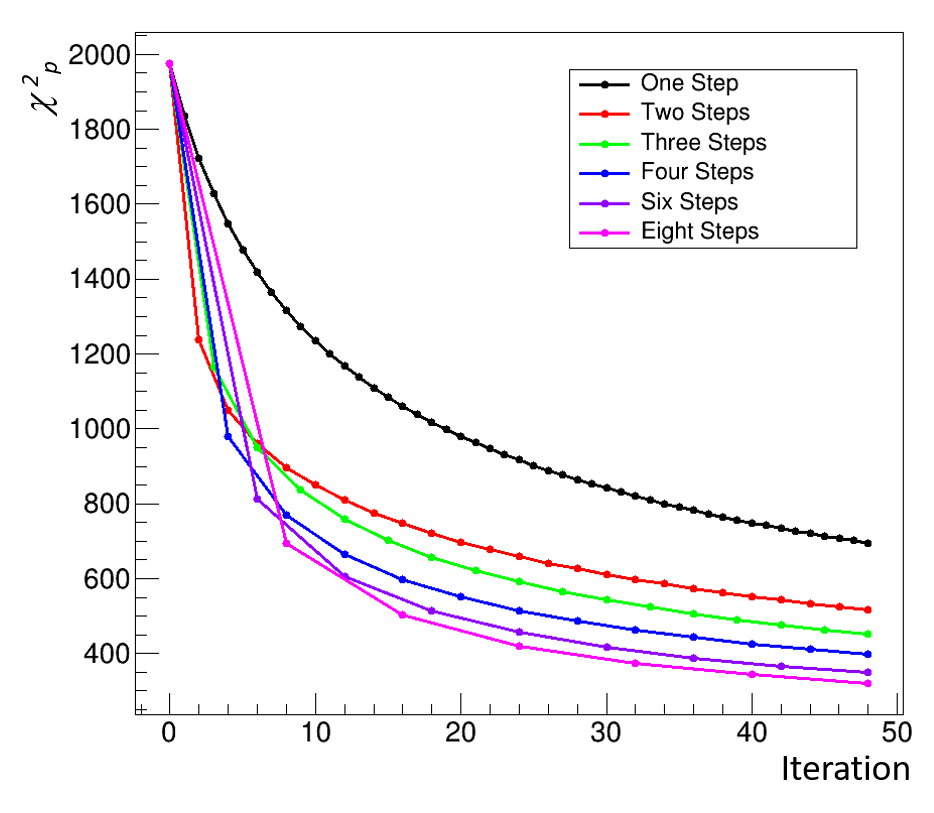}
  \caption{For a typical data set, $\chi^2_p$ versus iteration number, with simultaneous optimization of several steps, using (\ref{dpkn}).}\label{chi2p}
  \end{center}
\end{figure}

\begin{figure}
  \begin{center}
  \includegraphics[width=\columnwidth]{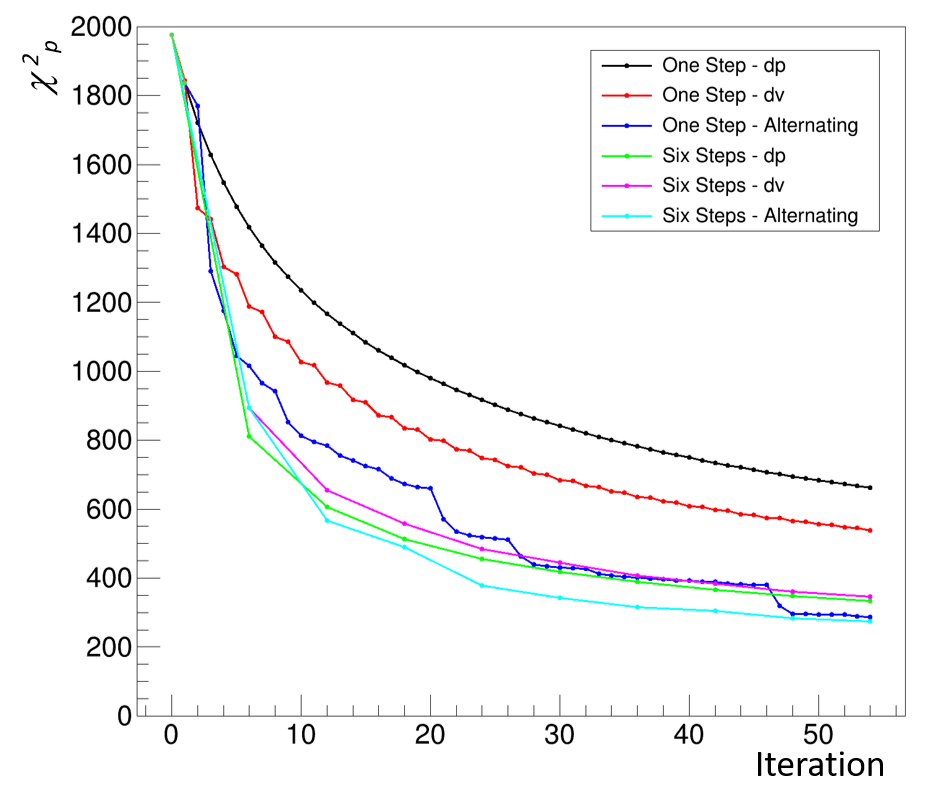}
  \caption{For a typical data set, $\chi^2_p$ versus iteration number for a variety of strategies, based on either (\ref{dpkn}) (dp), (\ref{dvkn}) (dv), or alternating between the two.}\label{chi2p_alt}
  \end{center}
\end{figure}

\begin{figure}
  \begin{center}
  \includegraphics[width=\columnwidth]{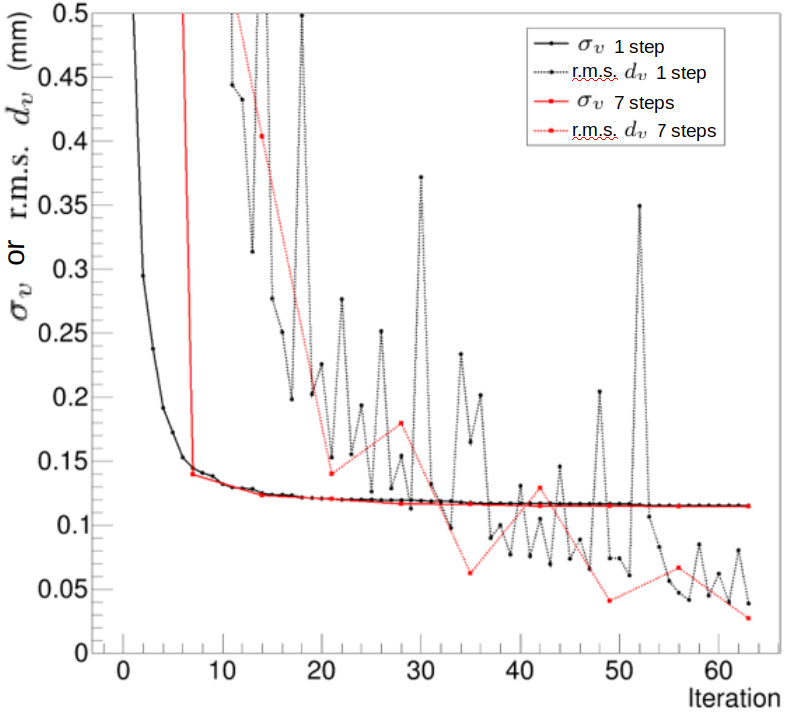}
  \caption{$\sigma_v$ as defined in (\ref{sigmav}), and r.m.s. $d_v$ versus iteration number, optimizing one step and seven steps using an alternating strategy, for the data used in Fig. \ref{george}. As r.m.s. $d_v$ falls below $\sigma_v$ the algorithm meets the stopping criterion in (\ref{stopping}) for common choices of $r$.}\label{sigma_v}
  \end{center}
\end{figure}

For our application, the entries for $A$ and $b$ are usually defined in mm, and the entries of $x$ have no units.  With repeated iterations, the {magnitudes of the resulting $p$ and $v$} vectors often increase rapidly.  In theory, this is not a problem, but in practice can affect the numerical stability of the solution of (\ref{dpkn}) or (\ref{dvkn}).  We resolve this by using for our units a length scale that maintains roughly constant magnitudes {of these vectors}.  This can be found with a few trials after the iterations are finished but before solving (\ref{dpkn}) or (\ref{dvkn}).  The length of the {voxel} volume is a good first guess in our experience. 

As illustrated in Fig.~\ref{chi2p}, we have found that optimizing many iterations simultaneously has major benefits in terms of the number of iterations needed to reduce $\chi^2$ to a given level. Fig.~\ref{chi2p_alt} compares optimizing based on $\chi^2_p$, $\chi^2_v$, or alternating between the two.  We have found that an alternating strategy provides the best convergence.  As discussed in Section~\ref{relaxation}, the steps based on $\chi^2_v$ bring the average $d_v$ closer to 0, and set up conditions for improved steps based on $\chi^2_p$. Steps based on $\chi^2_p$ often move the average $d_v$ away from 0.

The alternating strategy is powerful enough that a single step alternating strategy is often as optimal as a multi-step strategy. Fig.~\ref{sigma_v} compares the use of alternating strategies for both single step and seven step optimization for the data used for Fig.~\ref{george}, and shows similar performance for this example.

\section {Stopping Criteria}

The above methods optimize the $\chi^2$ of the solution after a number of iterations, and this $\chi^2$ can then be used to evaluate whether further iterations are needed or if the current solution is close enough to the optimal solution.  For example, worker processes can be continuously producing additional iterations of the $p$ and $v$ vectors while a parallel executive process finds the optimal coefficients and evaluates the quality of the fit.

We define a $\chi^2$ per degree of freedom, for which the square root can be interpreted as the average deviation per proton, as:
\begin{equation}
    \sigma_p = \sqrt{\frac{\chi^2_p}{N_p - N_v}}
\end{equation}
where $N_p$ is the total number of protons and $N_v$ is the total number of voxels.
With $N_{pv}$ as the average number of protons touching a voxel, as obtained from the data, we can define an estimated average voxel precision as:
\begin{equation}\label{sigmav}
    \sigma_v = \frac{\sigma_p}{\bar{\alpha}\sqrt{N_{pv}}}
\end{equation}
where $\bar{\alpha}$ is the average chord length of a proton through a voxel.  For our purposes, we can simply approximate this as the length of the side of a voxel.
If a region of the image is known to have uniform RSP, the estimated voxel precision can be determined from the image in that region, but in general the estimate in (\ref{sigmav}) has the advantage of not requiring assumptions about the RSP distribution.

At the minimum $\chi^2$, we expect $\sigma_p \approx$ 3 mm, based on our WEPL precision per proton, although in practice it tends to be somewhat larger, especially if the image has many non-uniform regions or sharp boundaries.  Since $d_v = 0$ at the minimum, if for a given iteration the root-mean-square (r.m.s) of $d_v$ is less than the estimated average voxel precision, it may be justified to stop iterating, since the noise from the proton measurement uncertainties is greater than the remaining distance to the optimal solution.
One example of a criterion to use in the decision to stop is:
\begin{equation}\label{stopping}
    \text{r.m.s.}\ d_v < r \sigma_v.  
\end{equation}
We typically choose $r$ in the range of 0.2 to 0.5, and we have found that with enough iterations we can generally reduce r.m.s. $d_v$ to any level.  Fig.~\ref{sigma_v} illustrates the evolution of r.m.s. $d_v$ and $\sigma_v$ with the number of iterations.  {It is also possible to use addition criteria in the decision to stop, such as a requirement on $\text{r.m.s.}\ d_v$ within a region of interest, or a requirement on the maximum outlier in the $d_v$ vector.}

\section{Strategies for memory use}

Memory resources can be a bottleneck in the implementation of these strategies.  While the $A$ matrix is very large, it is also very sparse, and various strategies to store the information in compact form are possible.  {One option is to recompute the elements of $A$ as they are needed, and this may be the best strategy if GPU processing is available}. As another example, the entries in $A$ can be stored as lists of voxels with chord lengths for each proton, or as lists of protons with chord lengths for each voxel.  Although the lists only include non-zero chord lengths, it still amounts to a large storage requirement.

In the case where we are storing the entries for $A$ as lists of voxels with chord lengths for each proton, we can take advantage of geometry to store this information with much less memory.  For example, each proton can have a list of line segments which can be used to recreate the voxel list and chord lengths when needed.  Each line segment should be short enough that a straight line approximates the proton trajectory to appropriate accuracy within the segment.

As another example, each proton can have a list of chord lengths, each typically stored in one byte, and a second list of base-6 numbers, stored in 4-byte integers with 12 base-6 numbers contained in each 4-byte integer.  Starting from a given voxel, the first base-6 number specifies the voxel face that the proton exits, and thus the identity of the next voxel, and subsequent base-6 numbers continue the chain from there.  Thus, the chord lengths can be associated with the correct voxel.

\section{Strategies for parallel processing}

The algorithms described in Ref. \cite{Penfold}, such as Diagonally Relaxed Orthogonal Projections (DROP), can process blocks of protons in parallel.  Various strategies combine the results from the different blocks. For example \cite{niu_patent}, worker processes find a solution for each block of data, a foreman process receives all the solutions, combines them, and sends the combined solution back to the worker processes for a further iteration.  One drawback of these approaches is that each block of protons must be large enough to solve the image, and the combined solution is not identical to what would be obtained with a single block.

{The least squares method makes possible}
 parallelization strategies that can use blocks with arbitrarily small numbers of protons and obtain a result which is exactly the same as if the calculations were executed in a single block.
Bottlenecks may involve memory resources, CPU resources, GPU resources, and data transfer capacity.  Choice of strategy will depend on the resources of a particular computing system, and implementation will rely on appropriate design of data structures and software architectures.

The computationally costly part of each full iteration involves a sequence of two matrix-vector multiplications for either $\bar{A}^T(Ad_{vk})$ or $\bar{A}^T(Av_k)$ as described above.  In a matrix-vector multiplication, each row of the matrix multiplies the vector independently, so it is possible to do all the row-vector multiplications in parallel, a task well suited to GPUs.

\subsection{Iteration with proton blocks followed by voxel blocks}
The most straightforward parallelization strategy is to:
\begin{enumerate}
    \item Divide the data into blocks of protons for the first multiplication, $Av_k$, processing each block in parallel.  The blocks may be as small as a single proton.
    \item Assemble the output vector, concatenating the output from each block.
    \item Divide the data into blocks of voxels for the second multiplication, $\bar{A}^T(Av_k)$, processing each block in parallel.  The blocks may be as small as a single voxel.
    \item Assemble the output vector, concatenating the output from each block.
\end{enumerate}
Parallel worker processes can execute the multiplications for each block of protons, a foreman process can assemble the outputs, and an executive process can evaluate the results as described above.  Blocks can be further divided into sub-blocks and a hierarchical system of blocks may help route calculations to multiple GPUs.

The drawback to this method is that the memory requirements exceed resources available in most currently cost-effective systems.  This method requires lists of voxels with chord lengths for each proton, as well as lists of protons with chord lengths for each voxel.  Although the lists only include non-zero chord lengths, it still amounts to a large storage requirement.  However, the cost of memory continues to drop, and this method may become feasible in the near future.

\begin{figure}
  \begin{center}
  \includegraphics[width=\columnwidth]{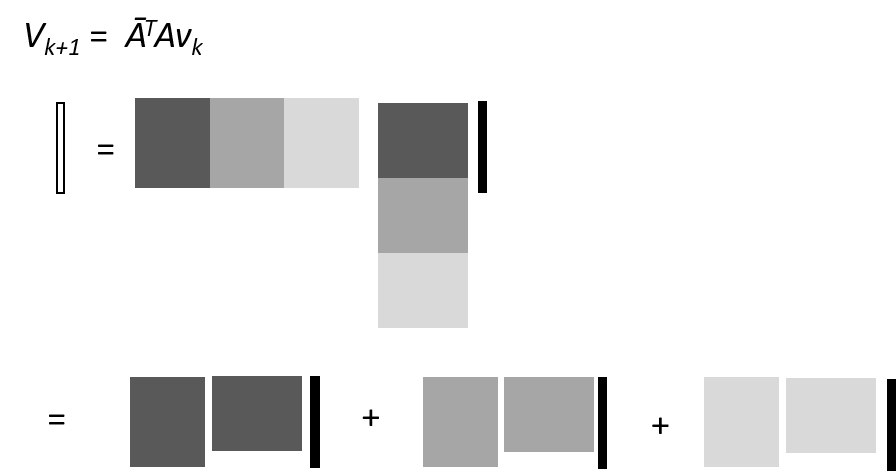}\\
  \caption{Illustration of the matrix-vector multiplications, showing chord length data for different blocks of protons with different shading. The chord length data for one proton is contained in one row of $A$, and a block consists of one or more rows. An iteration sequence can execute the two matrix-vector multiplications in blocks, as shown in the bottom line, with the additional step that a foreman process must concatenate the outputs from the first matrix-vector multiplication to create the $p_{k+1}$ vector, and sum the outputs of the second matrix-vector multiplication to create the $v_{k+1}$ vector.  However, the final result is the same as if the sequence were executed in a single block as shown in the upper line.}\label{blocks}
  \end{center}
\end{figure}

\subsection{Iteration with coordinated blocks of protons}
As illustrated in Fig.~\ref{blocks}, the entire sequence of multiplications may be carried out independently in different blocks of protons, each with a worker process, and with each block $t$ producing output vectors $[Av_k]_t$ and $[\bar{A}^T(Av_k)]_t$.  The foreman process concatenates the $[Av_k]_t$ vectors into the complete $p_{k+1}$ vector, and obtains the complete $v_{k+1}$ vector with a sum over the blocks:
\begin{equation}
    v_{k+1} = \sum_t [\bar{A}^T(Av_k)]_t
\end{equation}
The complete vectors are identical to what would be obtained with a single block and are the input to the next iteration.
Again, there can be a hierarchy of blocks.  If the computing system contains multiple GPUs, the data can be divided into one block for each GPU, {and further divided into sub-blocks after that. As shown in}~\cite{Hansen_2014}{for a similar algorithm, each proton can be processed as a separate block} utilizing the parallel processing power of the GPUs, carefully managing the summation of the output vectors from each proton.

Processing each proton independently (blocks of size one proton) enables major savings in the use of memory.  First, there is no need for a list of protons with chord lengths for each voxel.  All the needed information for the block is with the list of voxels with chord lengths for the proton, and the output vector needs only the voxels from that proton.  Second, we can take advantage of the path of the proton through adjacent voxels to store the list of voxels with much less memory, as described above.

\subsection{Iteration with independent blocks of protons}

This last method does not produce exactly the same results as for a single block but can be simply implemented without developing parallel processing architecture features such as foreman and executive processes, and provides a convenient method for rapid studies.  We start by dividing protons into $N$ well-randomized and equal-sized blocks $t$ (the method is trivially extendable to different sized blocks, as long as the assignment of a proton to a block is random).  Each block must contain enough protons to find a solution vector (more protons than voxels.) A program able to handle a single block can then be run with $N$ copies in parallel, each copy producing a solution close to the minimum $\chi^2$ for its block of data.

\begin{figure}
  \begin{center}
  \includegraphics[width=\columnwidth]{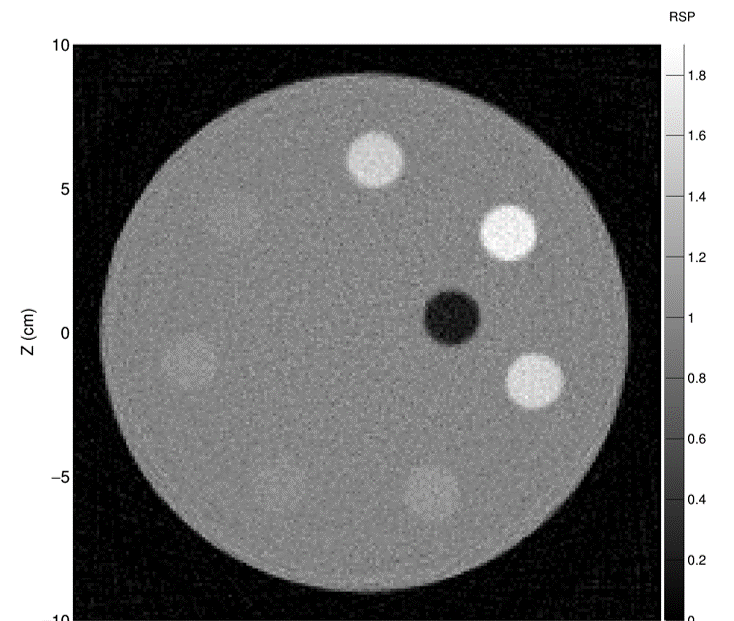}
  \includegraphics[width=\columnwidth]{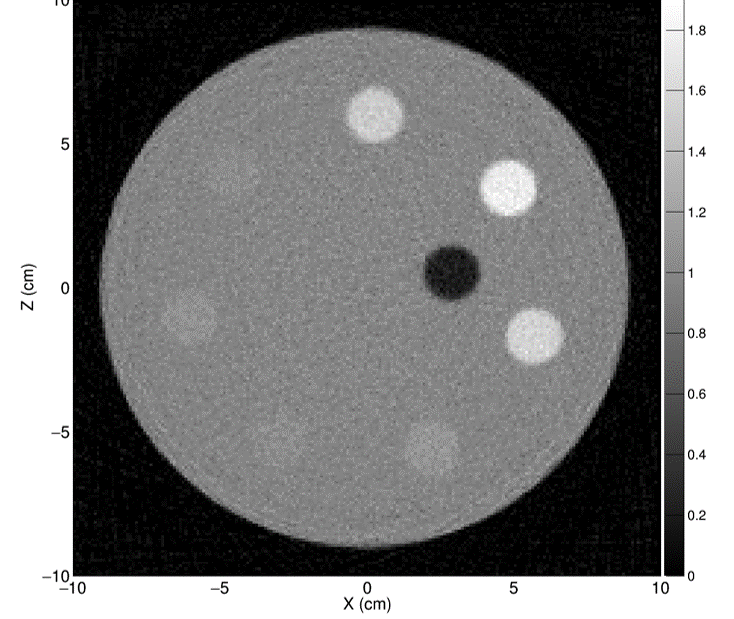}\\
  \caption{Top:  A 1 mm thick pCT slice of a cylindrical phantom with inserts, processed using a single block \cite{dejongh2020comparison}. Bottom:  Same data processed in 10 independent blocks.}\label{george}
  \end{center}
\end{figure}

Sums such as $\sum_j \alpha^T_{ij}$ scale linearly with the number of protons in the block. $\bar{A}^TA$ is a square matrix with dimension equal to the number of voxels where each entry is a ratio where the numerator and denominator both on average scale linearly with the number of protons.  Therefore, each entry is on average independent of the number of protons.  The following holds within the statistical variability of the data for each block, where the normalization of $\bar{A}^t$ is done using only the protons in that block:
\begin{gather}
    \bar{A}^T A \approx \bar{A}^T_t A_t \\
    V^{-1}_t \approx N V^{-1} 
\end{gather}
and we can show, at minimum $\chi^2$,
\begin{align}
\bar{A}^TA x &= \bar{A}^T b\\
    &= V^{-1} A^T b 
         = \sum_t V^{-1} A_t^T b_t \\	
              &= \sum_t V^{-1} V_t \bar{A}_t^T b_t 
                \approx \frac{1}{N} \sum_t \bar{A}_t^T A_t x\\
                &\approx   \bar{A}^TA \frac{1}{N} \sum_t x_t
\end{align}
\begin{equation}
          x  \approx \frac{1}{N} \sum_t x_t 
\end{equation}
and we find we can simply average the solution vectors from the blocks.
We have found this method works quite well, as shown for example in Fig.~\ref{george}, although with a little more noise near the edges with rapid RSP variation.  The cylindrical phantom in Fig.~\ref{george} contained inserts with known RSP, which are in good agreement with our measured values \cite{dejongh2020comparison}.

\section{Conclusion}

Clinically useful proton Computed Tomography images will rely on algorithms to find the three-dimensional proton stopping power distribution that optimally fits the measured proton data. The assurance of an image reconstruction quantifiably close to an optimal solution is a crucial step towards applying proton imaging technology to clinical treatment planning, and a useful starting point for evaluating the clinical impact of further image processing or use of approximations.  {While previous studies have generally analyzed reconstructed pCT images with respect to a known ground truth, we have developed metrics that are applicable to general unknown objects being imaged. }

While recent work has focused mostly on the use of iterative projective algorithms, we have revisited the approach of Goitein using iterative least squares algorithms.  We have found that this approach is suitable for proton imaging, which uses individual protons with relatively large measured WEPL uncertainties. Our method employs strategies with improved convergence which can naturally accommodate parallel processing and several features that help put pCT imaging into a  more quantitative framework.

\section*{Acknowledgment}

The authors thank our collaborators in the development of the pCT system:  V. Rykalin and I. Polnyi from ProtonVDA, M. Pankuch, B. Kreydick and G. DeFillippo from Northwestern Medicine Chicago Proton Center, N. Karonis, C. Ordoñez, K. Duffin, J. Winans, G. Coutrakon, C. Sarosiek and A. Best from Northern Illinois University, J. Welsh from Edward Hines Jr VA Hospital, and R. Schulte from Loma Linda University. 

% if have a single appendix:
%\appendix[Proof of the Zonklar Equations]
% or
%\appendix  % for no appendix heading
% do not use \section anymore after \appendix, only \section*
% is possibly needed

% use appendices with more than one appendix
% then use \section to start each appendix
% you must declare a \section before using any
% \subsection or using \label (\appendices by itself
% starts a section numbered zero.)
%

% ============================================
%\appendices
%\section{Proof of the First Zonklar Equation}
%Appendix one text goes here %\cite{Roberg2010}.

% you can choose not to have a title for an appendix
% if you want by leaving the argument blank
%\section{}
%Appendix two text goes here.

% use section* for acknowledgement
%\section*{Acknowledgment}

%The authors would like to thank D. Root for the loan of the SWAP. The SWAP that can ONLY be usefull in Boulder...

% Can use something like this to put references on a page
% by themselves when using endfloat and the captionsoff option.
\ifCLASSOPTIONcaptionsoff
  \newpage
\fi

% trigger a \newpage just before the given reference
% number - used to balance the columns on the last page
% adjust value as needed - may need to be readjusted if
% the document is modified later
%\IEEEtriggeratref{8}
% The "triggered" command can be changed if desired:
%\IEEEtriggercmd{\enlargethispage{-5in}}

% ====== REFERENCE SECTION

%\begin{thebibliography}{1}

% IEEEabrv,

\bibliographystyle{IEEEtran}
\bibliography{IEEEabrv,Bibliography}

\begin{thebibliography}{10}
\providecommand{\url}[1]{#1}
\csname url@rmstyle\endcsname
\providecommand{\newblock}{\relax}
\providecommand{\bibinfo}[2]{#2}
\providecommand\BIBentrySTDinterwordspacing{\spaceskip=0pt\relax}
\providecommand\BIBentryALTinterwordstretchfactor{4}
\providecommand\BIBentryALTinterwordspacing{\spaceskip=\fontdimen2\font plus
\BIBentryALTinterwordstretchfactor\fontdimen3\font minus
  \fontdimen4\font\relax}
\providecommand\BIBforeignlanguage[2]{{%
\expandafter\ifx\csname l@#1\endcsname\relax
\typeout{** WARNING: IEEEtran.bst: No hyphenation pattern has been}%
\typeout{** loaded for the language `#1'. Using the pattern for}%
\typeout{** the default language instead.}%
\else
\language=\csname l@#1\endcsname
\fi
#2}}
\renewcommand\BIBentryALTinterwordstretchfactor{4}

\bibitem{Scholz}
M.~{Scholz}, ``State-of-the-art and future prospects of ion beam therapy:
  Physical and radiobiological aspects,'' \emph{IEEE Transactions on Radiation
  and Plasma Medical Sciences}, vol.~4, no.~2, pp. 147--160, March 2020.

\bibitem{Lomax}
\BIBentryALTinterwordspacing
A.~J. Lomax, ``Myths and realities of range uncertainty,'' \emph{The British
  Journal of Radiology}, vol.~93, no. 1107, p. 20190582, 2020, pMID: 31778317.
  [Online]. Available: \url{https://doi.org/10.1259/bjr.20190582}
\BIBentrySTDinterwordspacing

\bibitem{Schreuder}
\BIBentryALTinterwordspacing
A.~N. Schreuder and J.~Shamblin, ``Proton therapy delivery: what is needed in
  the next ten years?'' \emph{The British Journal of Radiology}, vol.~93, no.
  1107, p. 20190359, 2020, pMID: 31692372. [Online]. Available:
  \url{https://doi.org/10.1259/bjr.20190359}
\BIBentrySTDinterwordspacing

\bibitem{Miller}
C.~Miller, B.~Altoos, E.~DeJongh, M.~Pankuch, D.~DeJongh, V.~Rykalin,
  C.~Ordoñez, N.~Karonis, J.~Winans, G.~Coutrakon, and J.~Welsh,
  ``Reconstructed and real proton radiographs for image-guidance in proton beam
  therapy,'' \emph{Journal of Radiation Oncology}, vol.~8, pp. 97--101, 03
  2019.

\bibitem{Ordonez}
C.~Ordoñez, N.~Karonis, K.~Duffin, J.~Winans, E.~DeJongh, D.~DeJongh,
  G.~Coutrakon, N.~Myers, M.~Pankuch, and J.~Welsh, ``Fast in situ image
  reconstruction for proton radiography,'' \emph{Journal of Radiation
  Oncology}, 05 2019.

\bibitem{Pankuch}
M.~Pankuch, E.~DeJongh, F.~DeJongh, and et~al., ``A method to evaluate the
  clinical utility of proton radiography for geometric patient alignment.'' in
  \emph{Proceedings of the 57th Annual Meeting of the Particle Therapy
  Cooperative Group (PTCOG)}, May 2018, available:
  http://theijpt.org/doi/pdf/10.14338/2331-5180-5-2-000.

\bibitem{Schulte}
\BIBentryALTinterwordspacing
R.~W. Schulte, V.~Bashkirov, M.~C. Loss~Klock, T.~Li, A.~J. Wroe, I.~Evseev,
  D.~C. Williams, and T.~Satogata, ``Density resolution of proton computed
  tomography,'' \emph{Medical Physics}, vol.~32, no.~4, pp. 1035--1046, 2005.
  [Online]. Available:
  \url{https://aapm.onlinelibrary.wiley.com/doi/abs/10.1118/1.1884906}
\BIBentrySTDinterwordspacing

\bibitem{Giacometti}
\BIBentryALTinterwordspacing
V.~Giacometti, V.~A. Bashkirov, P.~Piersimoni, S.~Guatelli, T.~E. Plautz,
  H.~F.-W. Sadrozinski, R.~P. Johnson, A.~Zatserklyaniy, T.~Tessonnier,
  K.~Parodi, A.~B. Rosenfeld, and R.~W. Schulte, ``Software platform for
  simulation of a prototype proton ct scanner,'' \emph{Medical Physics},
  vol.~44, no.~3, pp. 1002--1016, 2017. [Online]. Available:
  \url{https://aapm.onlinelibrary.wiley.com/doi/abs/10.1002/mp.12107}
\BIBentrySTDinterwordspacing

\bibitem{Rit}
\BIBentryALTinterwordspacing
S.~Rit, G.~Dedes, N.~Freud, D.~Sarrut, and J.~M. Létang, ``Filtered
  backprojection proton ct reconstruction along most likely paths,''
  \emph{Medical Physics}, vol.~40, no.~3, p. 031103, 2013. [Online]. Available:
  \url{https://aapm.onlinelibrary.wiley.com/doi/abs/10.1118/1.4789589}
\BIBentrySTDinterwordspacing

\bibitem{dejongh2020technical}
\BIBentryALTinterwordspacing
E.~A. DeJongh, D.~F. DeJongh, I.~Polnyi, V.~Rykalin, C.~Sarosiek, G.~Coutrakon,
  K.~L. Duffin, N.~T. Karonis, C.~E. Ordoñez, M.~Pankuch, J.~R. Winans, and
  J.~S. Welsh, ``Technical note: A fast and monolithic prototype clinical
  proton radiography system optimized for pencil beam scanning,'' \emph{Medical
  Physics}, vol.~48, no.~3, pp. 1356--1364, 2021. [Online]. Available:
  \url{https://aapm.onlinelibrary.wiley.com/doi/abs/10.1002/mp.14700}
\BIBentrySTDinterwordspacing

\bibitem{Johnson_2017}
\BIBentryALTinterwordspacing
R.~P. Johnson, ``Review of medical radiography and tomography with proton
  beams,'' \emph{Reports on Progress in Physics}, vol.~81, no.~1, p. 016701,
  nov 2017. [Online]. Available: \url{https://doi.org/10.1088/1361-6633/aa8b1d}
\BIBentrySTDinterwordspacing

\bibitem{sarosiek2020prototype}
\BIBentryALTinterwordspacing
C.~Sarosiek, E.~A. DeJongh, G.~Coutrakon, D.~F. DeJongh, K.~L. Duffin, N.~T.
  Karonis, C.~E. Ordoñez, M.~Pankuch, V.~Rykalin, J.~R. Winans, and J.~S.
  Welsh, ``Analysis of characteristics of images acquired with a prototype
  clinical proton radiography system,'' \emph{Medical Physics}, vol. Accepted
  Author Manuscript, 2021. [Online]. Available:
  \url{https://aapm.onlinelibrary.wiley.com/doi/abs/10.1002/mp.14801}
\BIBentrySTDinterwordspacing

\bibitem{dejongh2020comparison}
D.~F. DeJongh, E.~A. DeJongh, V.~Rykalin, G.~DeFillippo, M.~Pankuch, A.~W.
  Best, G.~Coutrakon, K.~L. Duffin, N.~T. Karonis, C.~E. Ordoñez, C.~Sarosiek,
  R.~W. Schulte, J.~R. Winans, A.~M. Block, C.~L. Hentz, and J.~S. Welsh, ``A
  comparison of proton stopping power measured with proton ct and x-ray ct in
  fresh post-mortem porcine structures,'' 2020, available:
  https://arxiv.org/abs/2012.06629.

\bibitem{Penfold}
S.~Penfold and Y.~Censor, ``Techniques in iterative proton ct image
  reconstruction,'' \emph{Sensing and Imaging}, vol.~16, 10 2015.

\bibitem{Schultze}
B.~Schultze, Y.~Censor, P.~Karbasi, K.~Schubert, and R.~Schulte, ``An improved
  method of total variation superiorization applied to reconstruction in proton
  computed tomography,'' \emph{IEEE Transactions on Medical Imaging}, vol.~PP,
  03 2018.

\bibitem{Goitein}
\BIBentryALTinterwordspacing
M.~Goitein, ``Three-dimensional density reconstruction from a series of
  two-dimensional projections,'' \emph{Nuclear Instruments and Methods}, vol.
  101, no.~3, pp. 509 -- 518, 1972. [Online]. Available:
  \url{http://www.sciencedirect.com/science/article/pii/0029554X72900390}
\BIBentrySTDinterwordspacing

\bibitem{Hansen_2016}
\BIBentryALTinterwordspacing
D.~C. Hansen, T.~S. S{\o}rensen, and S.~Rit, ``Fast reconstruction of low dose
  proton {CT} by sinogram interpolation,'' \emph{Physics in Medicine and
  Biology}, vol.~61, no.~15, pp. 5868--5882, jul 2016. [Online]. Available:
  \url{https://doi.org/10.1088/0031-9155/61/15/5868}
\BIBentrySTDinterwordspacing

\bibitem{Penfold2}
S.~N. {Penfold}, R.~W. {Schulte}, Y.~{Censor}, V.~{Bashkirov}, and A.~B.
  {Rosenfeld}, ``Characteristics of proton ct images reconstructed with
  filtered backprojection and iterative projection algorithms,'' in \emph{2009
  IEEE Nuclear Science Symposium Conference Record (NSS/MIC)}, 2009, pp.
  4176--4180.

\bibitem{Hansen_2014}
\BIBentryALTinterwordspacing
D.~C. Hansen, J.~B.~B. Petersen, N.~Bassler, and T.~S. SÃ¸rensen, ``Improved
  proton computed tomography by dual modality image reconstruction,''
  \emph{Medical Physics}, vol.~41, no.~3, p. 031904, 2014. [Online]. Available:
  \url{https://aapm.onlinelibrary.wiley.com/doi/abs/10.1118/1.4864239}
\BIBentrySTDinterwordspacing

\bibitem{Han}
G.~Han, G.~Qu, and Q.~Wang, ``Weighting algorithm and relaxation strategies of
  the landweber method for image reconstruction,'' \emph{Mathematical Problems
  in Engineering}, vol. 2018, pp. 1--19, 07 2018.

\bibitem{niu_patent}
N.~T. Karonis and et~al, ``High performance computing for three dimensional
  proton computed tomography (hpc-pct),'' U.S. Patent US8\,766\,180B2, May 2,
  2010.

\end{thebibliography}
%\end{thebibliography}
% biography section
% 
% If you have an EPS/PDF photo (graphicx package needed) extra braces are
% needed around the contents of the optional argument to biography to prevent
% the LaTeX parser from getting confused when it sees the complicated
% \includegraphics command within an optional argument. (You could create
% your own custom macro containing the \includegraphics command to make things
% simpler here.)
%\begin{biography}[{\includegraphics[width=1in,height=1.25in,clip,keepaspectratio]{mshell}}]{Michael Shell}
% or if you just want to reserve a space for a photo:

% ==== SWITCH OFF the BIO for submission
% ==== SWITCH OFF the BIO for submission
\vfill
\newpage
%\break

\begin{IEEEbiography}[{\includegraphics[width=1in,height=1.25in,clip,keepaspectratio]{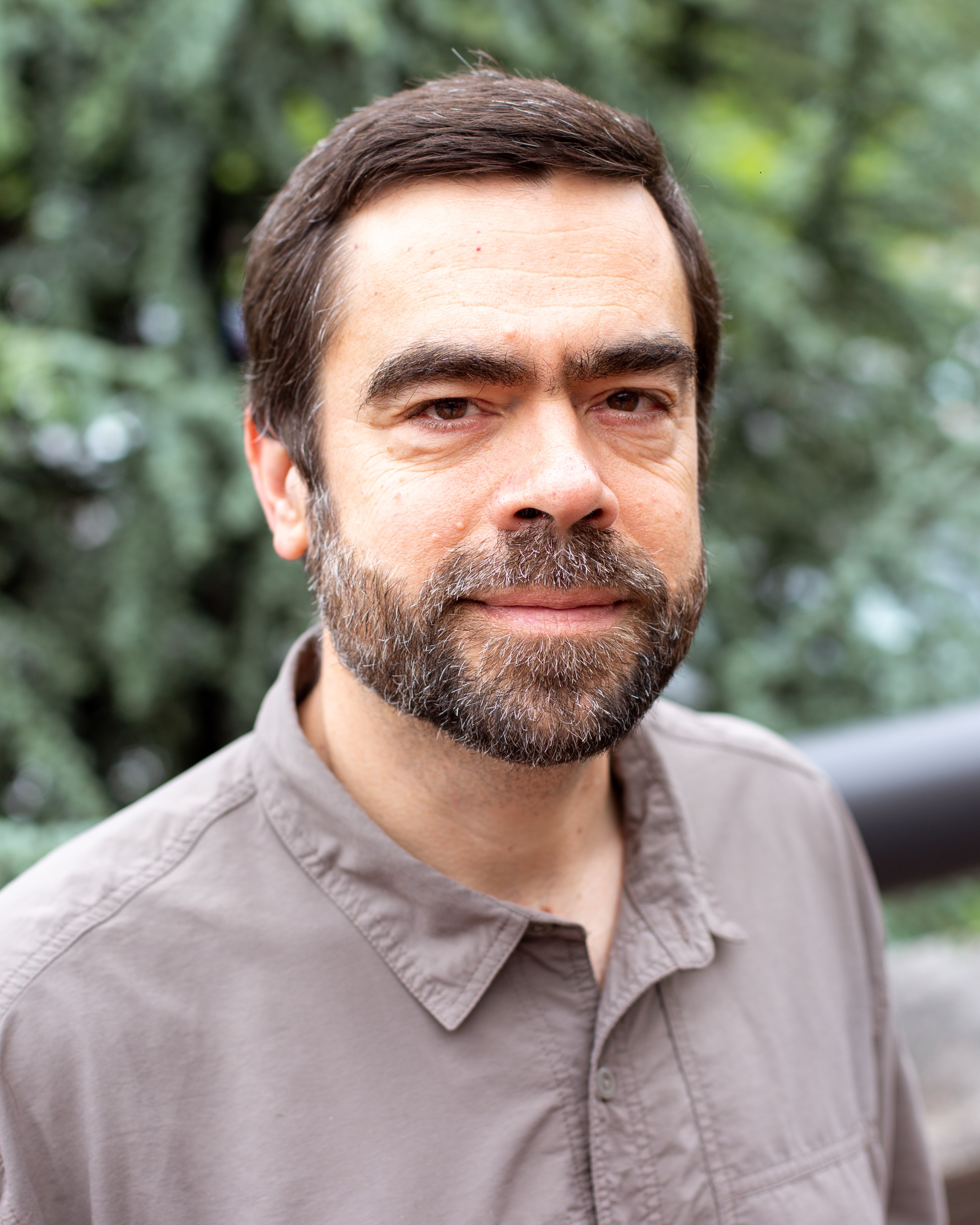}}]{Don "Fritz" DeJongh}
received the B.S. degree in physics and mathematics from the Ohio State University in 1983, and the Ph.D. degree in physics from the California Institute of Technology in 1990.  From 1990 to 2012, he was a Research Associate, Wilson Fellow, and Scientist at Fermilab, conducting research in particle physics and particle astrophysics. In 2014, he co-founded ProtonVDA, focusing on developing instrumentation to optimize proton radiation therapy.  He currently serves as CEO of ProtonVDA as well as Principle Investigator for the grant funding this work.
\end{IEEEbiography}
\begin{IEEEbiography}[{\includegraphics[width=1in,height=1.25in,clip,keepaspectratio]{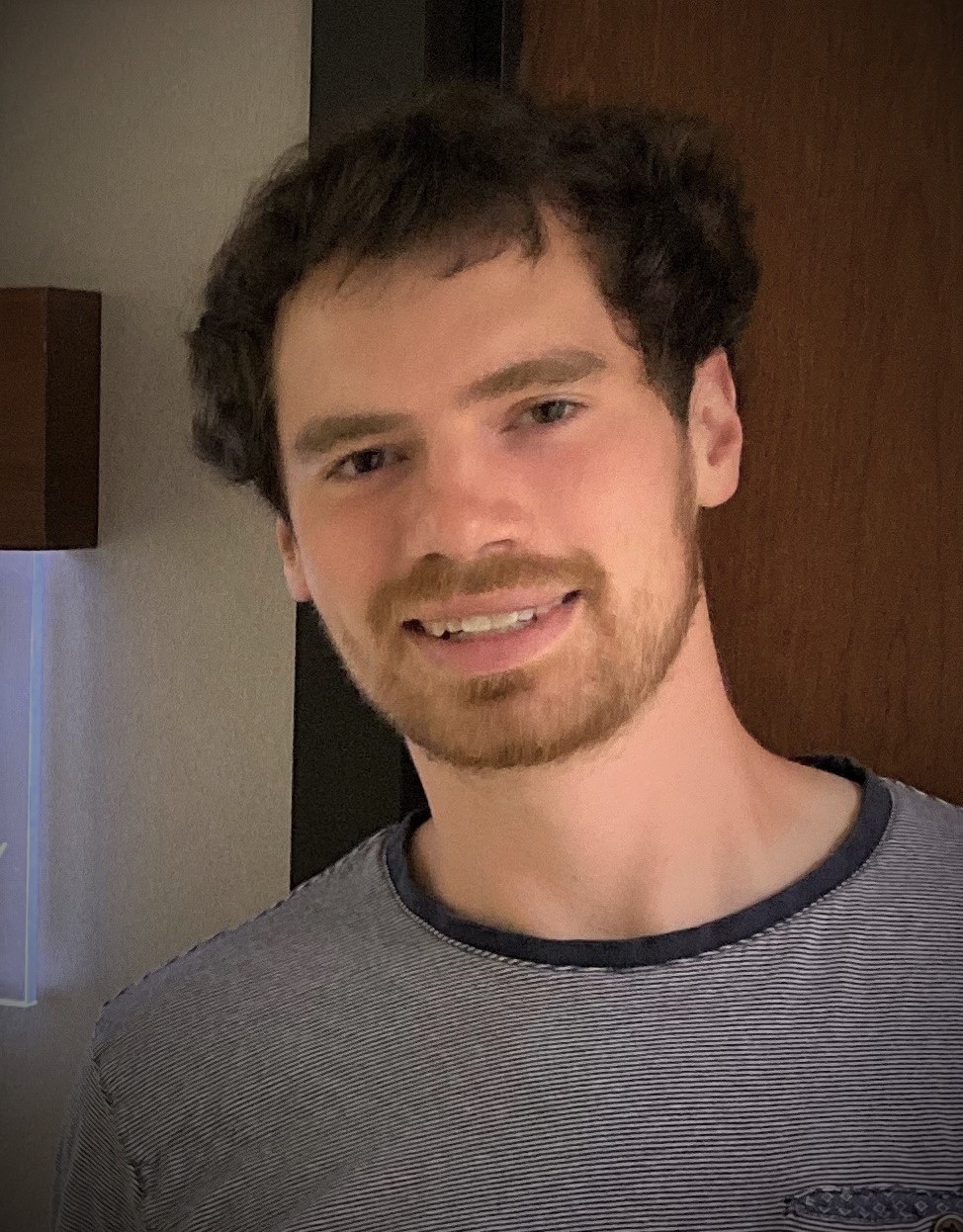}}]{Ethan A. DeJongh} received the B.S. degree in chemistry from the University of Chicago in 2014. He joined ProtonVDA as a research scientist in 2015, where he leads development of image reconstruction, simulations, and detector operations.
\end{IEEEbiography}

%% if you will not have a photo at all:
%\begin{IEEEbiographynophoto}{Ignacio Ramos}
%(S'12) received the B.S. degree in electrical engineering from the University of Illinois at Chicago in 2009, and is currently working toward the Ph.D. degree at the University of Colorado at Boulder. From 2009 to 2011, he was with the Power and Electronic Systems Department at Raytheon IDS, Sudbury, MA. His research interests include high-efficiency microwave power amplifiers, microwave DC/DC converters, radar systems, and wireless power transmission.
%\end{IEEEbiographynophoto}

%% insert where needed to balance the two columns on the last page with
%% biographies
%%\newpage

%\begin{IEEEbiographynophoto}{Jane Doe}
%Biography text here.
%\end{IEEEbiographynophoto}
% ==== SWITCH OFF the BIO for submission
% ==== SWITCH OFF the BIO for submission

% You can push biographies down or up by placing
% a \vfill before or after them. The appropriate
% use of \vfill depends on what kind of text is
% on the last page and whether or not the columns
% are being equalized.

\vfill

% Can be used to pull up biographies so that the bottom of the last one
% is flush with the other column.
%\enlargethispage{-5in}

% that's all folks
\end{document}